\documentclass[numberedappendix]{emulateapj}

\begin{document}

% document history:

\bibliographystyle{apj}

%\shorttitle{HR 8799 and HD 107146 Debris Disks}

%\shortauthors{Hughes et al.}

\slugcomment{Accepted for publication in ApJ: July 15, 2011}

\title{
Resolved Submillimeter Observations of the HR 8799 and HD 107146 Debris Disks
}

\author{A. Meredith Hughes\altaffilmark{1,2},
David J. Wilner\altaffilmark{3},
Sean M. Andrews\altaffilmark{3},
Jonathan P. Williams \altaffilmark{4},
Kate Y. L. Su\altaffilmark{5},
Ruth A. Murray-Clay\altaffilmark{3},
Chunhua Qi \altaffilmark{3}
}
\altaffiltext{1}{Department of Astronomy, University of California, Berkeley,
94720; mhughes$@$astro.berkeley.edu}
\altaffiltext{2}{Miller Fellow}
\altaffiltext{3}{Harvard-Smithsonian Center for Astrophysics, 60 Garden Street, Cambridge, MA 02138}
\altaffiltext{4}{Institute for Astronomy, University of Hawaii, 2680 Woodlawn
Dr., Honolulu, HI 96822}
\altaffiltext{5}{Steward Observatory, University of Arizona, 933 N Cherry Avenue, Tucson, AZ 85721}

\begin{abstract}

We present 880\,$\mu$m Submillimeter Array observations of the debris disks 
around the young solar analogue HD~107146 and the multiple-planet host star 
HR~8799, at an angular resolution of 3'' and 6'', respectively.  We spatially
resolve the inner edge of the disk around HR~8799 for the first time.  While 
the data are not sensitive enough (with rms noise of 1\,mJy) to constrain the 
system geometry, we demonstrate that a model by \citet{su09} based on the 
spectral energy distribution (SED) with an inner radius of 150\,AU 
predicts well the spatially resolved data.  Furthermore, by modeling 
simultaneously the SED and visibilities, we demonstrate that the dust is 
distributed in a broad (of order 100\,AU) annulus rather than a narrow ring.  
We also model the observed SED and visibilities for the HD~107146 debris disk 
and generate a model of the dust emission that extends in a broad band between 
50 and 170\,AU from the star.  We perform an {\it a posteriori} comparison 
with existing 1.3\,mm CARMA observations and demonstrate that a smooth, 
axisymmetric model reproduces well all of the available millimeter-wavelength 
data.  

\end{abstract}

\section{Introduction}

The tenuous, dusty debris disks orbiting main sequence stars are an important
tracer of planetary system architectures and their dynamical evolution.  With 
typical ages of 10-100\,Myr, debris disks are analogous to our own solar system 
during the period when giant and terrestrial planets were undergoing heavy 
bombardment by small bodies.  By this age, nearly all the primordial molecular 
gas has been dispersed from the protoplanetary disk, leaving a population of 
dust grains that should be removed by stellar radiation pressure on timescales 
much shorter than the age of the star.  The dust is therefore thought to be 
generated by the grinding together of planetesimals that formed from the 
primordial disk material.  Approximately 20\% of nearby main sequence stars 
exhibit detectable infrared excesses indicative of debris disks, implying that 
planet formation at least to the stage of planetesimal growth is a common 
process \citep[e.g.,][]{hab01,car05,bry06,moo10,mat11}.  However, despite 
their prevalence in the solar neighborhood, the relatively faint emission 
from orbiting dust grains makes imaging challenging, and only a handful of 
systems have thus far been spatially resolved.

%The dust in debris disks is generated and shaped by a combination of dynamical 
%processes involving planetesimals and remnant gas \citep[see][and references
%therein]{wya08}.  Spatially resolved observations have shown potential for 
%distinguishing between these processes, by detecting the dynamical signatures
%of unseen planets imprinted on the dust \citep[e.g.,][]{lio99,wil02} or 
%determining the ratio of small to large dust grains as a function of radius 
%by imaging at multiple wavelengths \citep[e.g.,][]{su05}.  Ample evidence has 
%accumulated to link features of debris disk structure -- including rings, 
%warps, clumps, and brightness asymmetries -- with the presence of planets in 
%the disk.  Two of the three planetary systems that have so far been directly 
%imaged were predicted from the geometry of the dust rings around the stars 
%\citep[Fomalhaut and $\beta$ Pictoris][]{kal08,lag10}.  Yet most debris disk 
%systems remain unresolved, with structural analyses relying on ambiguous 
%modeling of disk spectral energy distributions (SEDs). 

Observations at millimeter wavelengths have proven particularly valuable for 
studies of system dynamics, as these longer wavelengths are most sensitive to 
larger dust grains with long resonant lifetimes that trace best the structure 
resulting from orbital resonances \citep{wya06}.  
%The Vega debris disk, for 
%example, appears smooth and featureless at infrared wavelengths but at 
%millimeter wavelengths reveals clumpy structure likely due to the 
%gravitational influence of an orbiting planet \citep{hol98,wil02,su05,mar06,
%sib10}.  
To date, there have been only three debris disks spatially resolved
with interferometers: $\beta$ Pictoris, HD~107146, and HD~32297 
\citep{wil10,cor09,man08}.  In order to better understand the dust 
distribution in these systems and contribute to the sample of spatially 
resolved debris disk systems, we have observed two nearby, relatively bright 
debris disks with the Submillimeter Array (SMA), adding a new wavelength to 
the suite of observations of HD~107146 and spatially resolving the disk around 
HR~8799 for the first time.

HD~107146 is a rare example of a young, nearby, close solar analogue, with 
spectral type G2V \citep{jas78} and a distance of only 28.5\,pc from the Sun 
\citep{per97}.  Its age is estimated to lie between 80 and 200\,Myr, although 
its location on the H-R diagram allows for an age as young as 30\,Myr 
\citep{wil04}.  It was identified as an ``excess dwarf'' on the basis of 
its {\it IRAS} colors by \citet{sil00} and its debris disk was detected and 
marginally resolved for the first time at a wavelength of 450\,$\mu$m by 
\citet{wil04}.  It has since been spatially resolved in near-infrared 
scattered light \citep{ard04,ert11} as well as 350\,$\mu$m, 1.3\,mm and 
3\,mm dust continuum emission \citep{car05,cor09}.  Imaging 
reveals a broad debris belt with maximum brightness near a radius of 
$\sim$100\,AU from the star, although detailed modeling of the millimeter data
has not yet been carried out to determine its extent.  \citet{naj05} place an 
upper limit of 1 on the gas-to-dust mass ratio, indicating that the disk is 
largely devoid of primordial molecular gas.  \citet{cor09} found that their 
1.3\,mm image exhibits possible azimuthal brightness asymmetries, and 
suggested the presence of a planet that is shepherding the material into 
resonances.  Further observations of the large dust grain population are 
necessary to investigate that assertion.

HR~8799 is so far the only star to host a directly-imaged multiple-planet 
system.  Four planets have been observed with projected orbital radii ranging 
from $\sim$15-80\,AU \citep{mar08,mar10}.  HR~8799 also hosts a debris disk, 
and was one of the twelve originally identified ``Vega-like'' stars based on 
its 60\,$\mu$m {\it IRAS} flux \citep{sad86}.  SED modeling suggests that 
there are at least two dust belts in the system, likely bracketing the radii 
of the innermost and outermost planets known: the inner belt has a temperature 
of $\sim$150\,K and the outer belt has a temperature of $\sim$45\,K 
\citep{wil06,che09,su09}.  
%HR~8799 itself is both chemically and 
%asteroseismologically unusual: it is both a $\lambda$ Bootis star and a 
%$\gamma$ Doradus variable \citep{zer99}, but it is not a $\delta$ Scuti 
%pulsator despite its location within the classic $\delta$ Scuti instability 
%strip \citep{gra99}.  Like other $\lambda$ Bootis stars, it displays a metal 
%deficiency in Fe-peak elements of [Fe/H]=-0.47 \citep{gra99}, but a 
%roughly solar abundance of other elements \citep[including C and O;][]{sad06}. 
%While typically referred to in the literature as an AV5 star, \citet{gra99} 
%have assigned it the more detailed but arcane classification of kA5 hF0 mA5 v 
%$\lambda$ Boo.  
Like all isolated young A stars \citep[spectral type AV5, see e.g.][]{gra99}, 
its age is extremely difficult to determine, which is unfortunate because the 
masses of its planetary companions are estimated from hot-start models that 
depend sensitively on the assumed time since formation.  \citet{mar08} 
estimate an age between 30 and 160\,Myr based on several different indicators,
which implies planet-mass companions \citep[see also discussion in][]{hin10}. 
%like its galactic space motion, 
%position on the HR diagram, presence of a debris disk, and status as a 
%$\lambda$ Boo star; 
\citet{moy10} favor an age as old as 1\,Gyr based on an asteroseismological 
analysis, which would imply that the companions are in fact brown dwarfs, 
alothough an infrared color analysis by \citet{cur11} argues against such a 
high mass for the companions.  The presence of a molecular gas filament 
coincident in space and recession velocity with HR~8799 \citep{wil06,su09} may 
also point to a younger age.  Spatially resolving the disk is therefore 
crucial, because it can provide new constraints on the companion masses.  Both 
the age determination and the stability of the planetary system depend 
sensitively on the unknown viewing geometry \citep{fab10}, which can be 
determined by spatially resolved observations of the disk, assuming 
coplanarity.  Spatially resolving the dust distribution, particularly at long 
wavelengths, also places a more direct dynamical constraint on the mass of 
the outermost planet independent of the age of the system \citep[see, 
e.g.,][]{chi09}.  

To investigate the spatial distribution of large dust grains and their 
implications for inferring the presence and properties of planets in the 
systems, we observed HD~107146 and HR~8799 with the SMA at a frequency of 
880\,$\mu$m.  The observations are described in Section~\ref{sec:obs} and the 
results are presented in Section~\ref{sec:results}.  We generate a model of 
the disk around HD~107146 that is capable of reproducing both the SED and SMA 
visibilities and compare it with the previous 1.3\,mm CARMA observations in 
Section~\ref{sec:hd_analysis}.  We perform a simplified analysis on
the lower signal-to-noise HR~8799 data in Section~\ref{sec:hr_analysis}, 
comparing the data with the SED-based model from \citet{su09} and
investigating whether the data can distinguish between a broad and narrow 
dust ring.  We discuss the implications of these results for the properties 
of the systems in Section~\ref{sec:discussion}, and summarize our conclusions
in Section~\ref{sec:summary}.

\section{Observations}
\label{sec:obs}

HR~8799 was observed during four nights in the fall of 2009.  Basic observing
parameters are listed in Table~\ref{tab:obs}.  Precipitable water 
vapor was very low on all three nights, with the 225\,GHz opacity holding 
steady at around 0.06.  Phase stability was more variable: September 20 had 
the best phase stability, while the second half of the night on October 1 was 
nearly unusable, exhibiting substantial phase decorrelation corresponding to 
large spikes in humidity measured at the summit of Mauna Kea.  Observations
of HR 8799 were interleaved with 3C454.3, one of the brightest quasars in the
sky and only 6$^\circ$ away from the target, to calibrate the atmospheric and 
instrumental phase variations.  The quasar J2232+117 was included in the 
observing loop to test the efficacy of the phase transfer.  To emphasize the 
emission on the shortest baselines, we image the combined dataset with a 6" 
taper, resulting in an rms noise of 1.1\,mJy\,beam$^{-1}$. 

HD~107146 was observed during four nights in early 2009.  Basic observing 
parameters are listed in Table~\ref{tab:obs}.  The weather was excellent 
on all nights, although the night of January 21 was particularly spectacular 
with a 225\,GHz opacity dropping rapidly from 0.06 to 0.03, and remaining low 
throughout the night.  Observations of HD~107146 were interleaved with the 
quasars 3C273 (15$^\circ$ away) and 3C274 (5$^\circ$ away) to calibrate the 
atmospheric and instrumental gains.  Despite the presence of a bright, 
spatially extended jet observed at longer radio wavelengths \citep[e.g.][]{
owe80}, we detect extended structure around 3c274 at a level of only a few 
percent of the core flux, at a position angle coincident with the known jet 
location.  The marginally resolved 3c274 structure had minimal effect on the 
resulting visibilities, with phase transfer to 3c273 resulting in point-like 
visibilities to within $<$0\farcs3.  We image the combined dataset with a 
1\farcs2 taper, resulting in an rms noise of 0.7\,mJy\,beam$^{-1}$.

The local oscillator (LO) frequency was 341.493\,GHz for the HR~8799 
observations and 340.755\,GHz for the HD~107146 observations.  The full 8\,GHz 
bandwidth (4\,GHz in each sideband separated by $\pm$5\,GHz from the LO 
frequency) was sampled evenly at a relatively low spectral resolution of 
0.7\,km\,s$^{-1}$ to maximize continuum sensitivity.  Routine calibration 
tasks were carried out using the MIR\footnote{See 
http://cfa-www.harvard.edu/$\sim$cqi/mircook.html} software package, while 
imaging and deconvolution were accomplished using the MIRIAD software package.
To estimate the effects of phase instability on seeing, we fit Gaussian
functions to the test quasars and measure sizes of 0\farcs1$\pm$0\farcs1 for
3C273 (HD 107146) and 0\farcs2$\pm$0\farcs3 for J2232+117 (HR 8799).  The 
3$\sigma$ upper limits on the seeing are approximately an order of magnitude 
lower than the spatial resolution of the final images.

\begin{table*}
\caption{Basic Observing Parameters}
\begin{tabular}{lcccc}
\hline
Date & Baseline lengths (m) & 225\,GHz opacity & Flux calibrator & Derived gain cal flux$^a$ (Jy) \\
\hline
\multicolumn{5}{c}{HR 8799} \\
\hline
Sep 20 & 9.5-68 & 0.06 & Ganymede & 16.1 \\
Sep 21 & 9.5-68 & 0.06 & Callisto & 16.7 \\
Oct 1 & 9.5-68 & 0.06 & Callisto & 19.9 \\
\hline
\multicolumn{5}{c}{HD 107146} \\
\hline
Jan 6 & 16-69 & 0.07 & Titan & 1.3 \\
Jan 21 & 9.5-69 & 0.03-0.06 & Titan & 1.2 \\
Feb 1 & 9.5-69 & 0.06 & Ceres & 1.1 \\
May 2 & 25-139 & 0.05 & Titan & 1.1 \\
\hline
\end{tabular}
\tablenotetext{a}{The quasars used as gain calibrators were 3C454.3 for HR 8799 and 3C274 for HD 107146.}
\label{tab:obs}
\end{table*}

%\begin{table}
%\caption{HR 8799 Observation Log}
%\begin{tabular}{lccc}
% & September 20 & September 21 & October 1 \\
%\hline
%Baseline lengths (m) & 9.5-68 & 9.5-68 & 9.5-68 \\
%225\,GHz opacity & 0.06 & 0.06 & 0.06 \\
%Flux calibrator & Ganymede & Callisto & Callisto \\
%Derived 3C454.3 Flux (Jy) & 16.1 & 16.7 & 19.9 \\
%\hline
%\end{tabular}
%\label{tab:hr8799}
%\end{table}
%
%\begin{table}
%\caption{HD 107146 Observation Log}
%\begin{tabular}{lcccc}
% & January 6 & January 21 & February 1 & May 2\\
%\hline
%Baseline lengths (m) & 16-69 & 9.5-69 & 9.5-69 & 25-139 \\
%225\,GHz opacity & 0.07 & 0.03-0.06 & 0.06 & 0.05 \\
%Flux calibrator & Titan & Titan & Ceres & Titan \\
%Derived 3C274 Flux (Jy) & 1.3 & 1.2 & 1.1 & 1.1 \\
%\hline
%\end{tabular}
%\label{tab:hd107146}
%\end{table}

\section{Results}
\label{sec:results}

\begin{figure*}
\epsscale{1.0}
\plotone{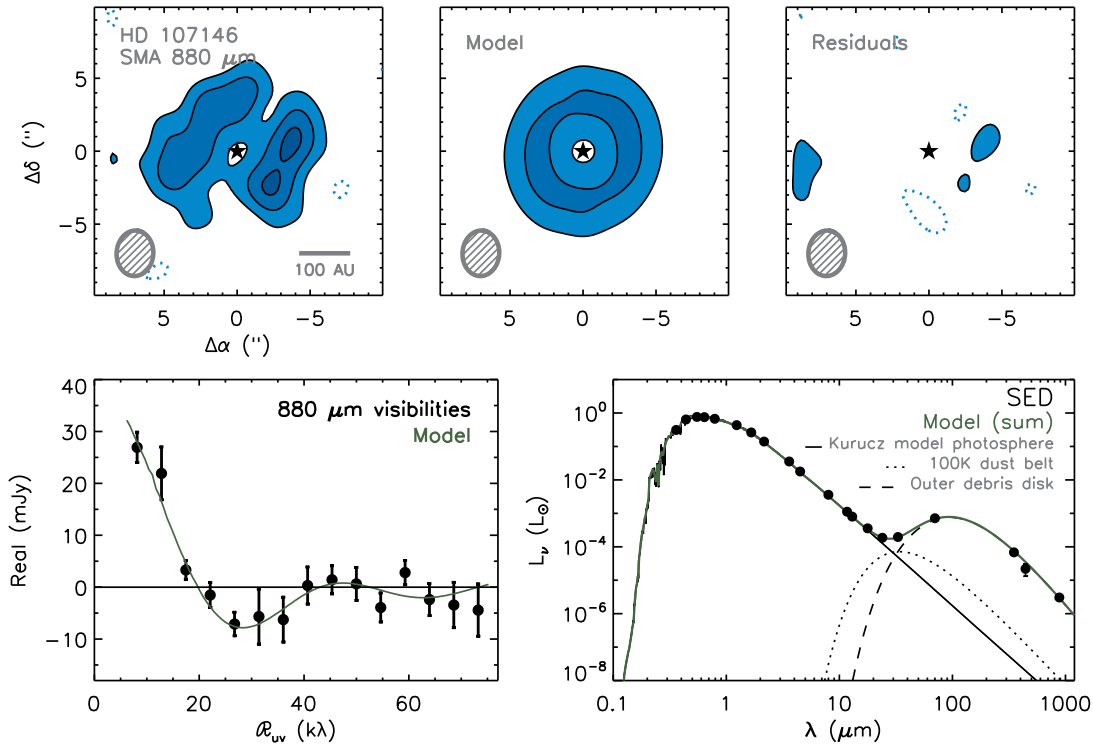}
\figcaption{Comparison between HD 107146 data and the best-fit model for the 
SED and visibilities.  The top row shows the SMA 880\,$\mu$m data (left), model
(center), and residuals (right) in the image domain, with contours 
[2,4,6]$\times$0.7\,mJy\,beam$^{-1}$ (the rms noise).  The 
2\farcs5$\times$3\farcs1 synthesized beam is indicated in the lower left of 
each panel.  The bottom row compares the data (black points) and model (green 
line) in the visibility domain (left) as well as the SED (right).  The model 
SED is the sum of three components: a Kurucz model photosphere (solid black 
line), a warm inner dust belt (dotted line), and a cold outer dust belt (dashed 
black line) that contributes effectively all the emission at millimeter 
wavelengths.  The units of the ordinate are defined so that $L_{\nu} = 4\pi
d^2 \nu F_{\nu}$ in units of $L_{\sun}$. 
\label{fig:hd107146}}
\end{figure*}

We detect 880\,$\mu$m continuum emission from HD~107146 on all four nights; the
combined map and visibilities are displayed in the left panels of 
Figure~\ref{fig:hd107146}.  The emission is distributed in a ring around the
position of the star, which is adjusted for proper motion and marked with a
star symbol in the figure.  The peak brightness of the ring occurs at 
approximately 3\farcs5 from the star position, which corresponds to a linear 
separation of 100\,AU at a distance of 28.5\,pc.  There is a deficit of 
emission coincident with the star position that suggests the presence of a
central cavity in the dust distribution.  Using the MIRIAD task {\tt cgcurs}, 
we estimate an integrated flux density of 36$\pm$1\,mJy.  

Due to periods of atmospheric phase instability, large portions of the nights 
of September 20 and October 1 were unusable, resulting in an rms noise for each
night that was approximately a factor of two larger than the rms noise on
September 21.  As a result, we detect emission from the HR~8799 disk only on 
the night of September 21.  The left panel of Figure~\ref{fig:hr8799} displays 
the combined map and the right panel shows the visibilities.  The peak 
signal-to-noise ratio is 4.3, and the emission peak occurs 5.5'' from the 
star.  There is some suggestion that the emission is slightly extended along 
an arc of the same radius, as would be expected if it were distributed in a 
ring around the star.  It is impossible to determine from the images alone 
the underlying radial distribution of the dust grains or the degree of 
azimuthal (a)symmetry of the emission, due to the low signal-to-noise ratio 
and incomplete sampling of the ($u$,$v$) plane.  In Section~\ref{sec:analysis}, 
we therefore model the underlying dust distribution to constrain its extent 
and to search for asymmetries in the residuals.

\subsection{Line-of-Sight CO Towards HR~8799}

We detect the CO(3-2) emission reported by \citet{wil06} and demonstrated by
\citet{su09} to be at an LSR velocity coincident with that of HR~8799.  
The line emission is quite narrow, appearing in only two of the 
0.7\,km\,s$^{-1}$ channels, indicating a linewidth of 
$\lesssim$1.4\,km\,s$^{-1}$.  The line does not image cleanly in the 
interferometric data, exhibiting striping from northeast to southwest across
the image.  There are two likely reasons for this: (1) the position angle of 
the stripes matches that of the CO filament in Figure 4 of \citet{wil06}, and
the separation between stripes is equal to that between the peak of the
synthesized beam and its largest (30\%) sidelobes, and (2) there is a knot of
bright CO emission visible in the JCMT map to the northwest of the SMA map 
center at a distance of approximately the width of the SMA primary beam, 
which coincides with a peak in the SMA map.  Not only are interferometers 
inherently poorly suited to imaging extended emission like the CO filament
observed with the JCMT, but the presence of a bright source outside the 
primary beam on its own would be expected to cause striping across the image.
Nevertheless, there is no evidence of a velocity gradient across the line, 
which is characteristic of rotation and would be expected if the CO(3-2) 
emission were associated with the disk.  Given the spatial resolution of the
SMA, such a gradient should be easily detectable if it were present.  Assuming
a gas disk spatially coincident with the observed dust emission, the CO(3-2)
line should be concentrated well within the 30'' SMA primary beam and should
exhibit centroid shifts across the line by $\sim$5''  between the red 
and blue channels, which is not observed.  We therefore conclude that the CO 
emission does not originate from the disk around HR~8799, although the 
coincidence in spatial location and velocity are suggestive that the disk 
may be associated with the CO filament.

%\begin{figure*}
%\epsscale{1.0}
%\plotone{hr8799_co.eps}
%\figcaption{
%\label{fig:hr8799_co}}
%\end{figure*}

\section{Analysis}
\label{sec:analysis}

In order to constrain the spatial distribution of circumstellar dust in these
systems, we model simultaneously the spectral energy distribution (SED) and 
spatially resolved 880\,$\mu$m visibilities.  Section~\ref{sec:hd_analysis} 
describes the modeling procedure used for HD~107146, including an {\it a 
posteriori} comparison between the model of the SMA data and previous 
spatially resolved observations of HD~107146 with CARMA.  
Section~\ref{sec:hr_analysis} describes how we modify this procedure for the 
low signal-to-noise case of HR~8799.  

\subsection{HD~107146 Modeling Procedure}
\label{sec:hd_analysis}

The bottom right panel of Figure~\ref{fig:hd107146} shows the HD 107146 SED
assembled from the literature \citep[][plus IRAS and 2MASS fluxes]{lan83,
wil04,car05,moo06,hil08,cor09}.  We model the SED with three components:
(1) a Kurucz model stellar photosphere with surface gravity $\log{g}$=4.5 and 
effective temperature $T_{eff}$=5859\,K \citep{car08,hil08}; (2) a dust belt 
with a temperature of 69\,K, which is required to reproduce the mid-IR fluxes 
but does not contribute substantially to the millimeter-wavelength flux; and 
(3) an outer debris belt with the properties described below.  It is this 
latter component that accounts for effectively all of the 880\,$\mu$m flux, 
and it must therefore be modeled as a spatially extended component to 
reproduce both the observed visibilities and the SED.

As in \citet{wil04}, we assume a dust grain emission efficiency $Q_\lambda = 
1 - \exp{[-(\lambda_0/\lambda)^{\beta}]}$, where $\lambda_0$ is a critical 
wavelength; this has the desired asymptotic behavior that $Q_\lambda = 
(\lambda_0/\lambda)^\beta$ for $\lambda >> \lambda_0$ and $Q_\lambda = 1$ 
for $\lambda << \lambda_0$, while varying smoothly between the two extremes.
However, unlike \citet{wil04} who fit the SED with a single dust temperature, 
we require the dust grains to be in radiative equilibrium with the star and 
allow the disk to be spatially extended.  With these assumptions, using the 
textbook formulation from \citet{tie05}, the dust temperature is given by
\begin{eqnarray}
T_r = \left[ \frac{L_*}{\sigma r^2} \frac{\pi^3}{240} 
\frac{1}{(\beta+3)! \zeta(\beta+4) Q_0} \left(\frac{\lambda_0 k}{hc}\right)^{-\beta}\right]^\frac{1}{4+\beta},
\end{eqnarray}
where $L_*$ is the stellar luminosity, $\sigma$ is the Stefan-Boltzmann 
constant, $r$ is the distance from the star, $\beta$ is the dust grain emission
efficiency power law index, $\zeta$ is the Riemann zeta function, and $Q_0$ is 
the dust grain efficiency at the critical wavelength $\lambda_0 = 2 \pi a$.  
We assume a single characteristic grain size $a$, since this is sufficient to 
model the data and more complexity in the grain size distribution is not 
required to fit currently available data.  
%The mass opacity $\kappa_\lambda$ 
%is given by $\frac{3 Q_\lambda}{4 a \rho_\mathrm{dust}}$, where $a$ is the 
%characteristic dust grain size and $\rho_\mathrm{dust}$ is the dust density, 
%assumed to be 2.7\,g\,cm$^{-3}$.  
The flux as a function of wavelength is then 
\begin{eqnarray} 
F_\lambda =\frac{\pi a^2 Q(\lambda)}{d^2} \int_{r_\mathrm{in}}^{r_\mathrm{out}} \! 2 \pi r
\, B_\lambda(T_r) \, n(r)\, \mathrm{d}r,
%F_\lambda =\frac{3}{4 a \rho d^2} Q(\lambda) \int_{r_\mathrm{in}}^{r_\mathrm{out}} \! 2 \pi r
%\, B_\lambda(T_r) \, \Sigma(r)\, \mathrm{d}r,
\end{eqnarray}
where $d$ is the distance to the source, $r_\mathrm{in}$ and $r_\mathrm{out}$ 
are the inner and outer radii of the dust disk, $B_\lambda$ is the Planck 
function, and $n$ is the number density of dust grains.  The number density of
grains is related to the surface mass density of dust grains as $\Sigma = n 
m_g$, where $m_g$ is the mass per dust grain, and we parameterize $\Sigma$ as 
a function of radius in the following way: $\Sigma(R) = \Sigma_{100}(100\,\mathrm{AU}/r)^p$.  For consistency with previous works \cite[e.g.,][]{wil04}, we
assume a standard millimeter-wavelength mass opacity of 
$1.8(\nu/10^{12}\,\mathrm{Hz})^{0.8}$\,cm$^{2}$\,g$^{-1}$ \citep{dal01} 
evaluated at the 880\,$\mu$m wavelength of the SMA data. 

Because the outer dust belt is not bright enough at short wavelengths to
reproduce the 24 and 33\,$\mu$m {\it Spitzer} MIPS fluxes, we add a low-mass 
69\,K dust belt component to the SED (it would be entirely unresolved and 
contribute no detectable flux in the SMA data).  The temperature is set to 
match that derived from the IR spectrum in \citet{car09}, and the mass is 
then varied to best reproduce the observed SED.  We do not expect the 
properties of this belt to substantially affect the properties of the outer
disk that is the focus of this analysis. 

We also generate a model image assuming an inclination $i$=25$^\circ$ and 
position angle of the major axis $PA$=148$^\circ$ from the scattered light 
images \citep{ard04}.  We compare the spatially resolved model to the SMA 
observations in the visibility domain, using the MIRIAD task \texttt{uvmodel} 
to sample the model image at the same spatial frequencies as the SMA data
after convolution with a 0.3" Gaussian seeing kernel.

To find the best-fit model parameters, we perform a $\chi^2$ comparison 
between the model and the SED and SMA visibilities.  We calculate a $\chi^2$ 
value for the SED as a whole as well as for all of the real and imaginary 
visibilities at each spatial frequency and then simply sum the two values so 
that $\chi^2 = \chi_\mathrm{SED}^2 + \chi_\mathrm{vis}^2$.  As discussed in 
\citet{and09}, due to a fortuitous balance between the quality and quantity 
of samples in the two data sets, each contributes roughly the same amount to 
the total $\chi^2$ value, so that neither the SED nor the visibility data 
dominates the final fit.  We vary the dust grain properties \{$a$,$\beta$\}, 
the spatial parameters \{$r_\mathrm{in}$,$r_\mathrm{out}$,$p$\}, and the total 
mass in the disk $M_\mathrm{disk}$ and in the belt $M_\mathrm{belt}$.  The 
total disk mass is related to the surface density as $M_\mathrm{disk} = 2\pi 
\Sigma_{100}(100\,AU)^p \frac{r_\mathrm{out}^{2-p} - r_\mathrm{in}^{2-p}}{2-p}$.  

The six parameters effectively divide into two sets of three, with the spatial
parameters \{$r_\mathrm{in}$,$r_\mathrm{out}$,$p$\} allowing us to reproduce
the morphology of the spatially resolved data and the dust parameters
\{$a$,$\beta$,$M_\mathrm{disk}$\} allowing us to match the SED.  The 
characteristic grain size $a$ affects the dust temperature and shifts the 
peak wavelength of the blackbody.  $\beta$ has a similar effect, but its 
influence on $Q_\lambda$ is far more visible in determining the overall 
shape of the dust spectrum, particularly the slope with which the tail of the 
curve falls at long wavelengths.  For the optically thin disk models we 
consider, the total mass $M_\mathrm{disk}$ acts as a normalizing factor, 
shifting the blackbody curve vertically to match the observed integrated 
millimeter flux.  The visibilities are particularly sensitive to the inner
radius $r_\mathrm{in}$, which is the dominant factor in determining the 
location of the first null in the visibility function, i.e., the shortest
baseline at which the visibilities pass through zero and become negative.
However, the narrower the ring, the greater the influence of the outer radius
$r_\mathrm{out}$ on the null location, which can also be further moderated by
how concentrated the emission is towards the inner radius, described by the
surface density power law $p$.  There is a fairly strong degeneracy between
$r_\mathrm{out}$ and $p$; however, we include both parameters because the 
surface density power law is of interest for comparisons with the 
shorter-wavelength data.

We perform the $\chi^2$ minimization in multiple steps, beginning with coarsely
sampled grids and increasing the grid resolution to refine our estimates and
avoid settling on a local rather than a global minimum.  The initial coarse grid
allows $a$ to vary between 0.01 and 100\,$\mu$m, $\beta$ between -1 and 1.5, 
$M_\mathrm{disk}$ between 10$^{-1}$ and 10$^{-4}$\,M$_\earth$, $r_\mathrm{in}$ 
and $r_\mathrm{out}$ between 1 and 500\,AU, and $p$ from -1 to 4.
Eventually the grids are refined so that the smallest (linear) steps are 
1\,$\mu$m for $a$, 0.1 for $\beta$, 10\,AU for $r_\mathrm{in}$ and 
$r_\mathrm{out}$, 10$^{-3}$\,M$_\earth$ for $M_\mathrm{disk}$, 0.1 for $p$, and
2$\times$10$^{-4}$\,M$_\earth$ for $M_\mathrm{belt}$.  The uncertainties on each
parameter are larger than the steps, although they are also frequently 
correlated with other parameters.  Given the low signal-to-noise ratio of
the data and the rapidly improving capabilities of millimeter interferometers,
our intention is not to provide a detailed characterization of the disk and a
robust statistical analysis of the errors, but rather to generate a 
representative model that can reproduce the gross features of the SED and 
visibilities (both SMA and CARMA, below) to constrain basic properties like 
the size, width, mass, and degree of axisymmetry of the dust ring.  

The best-fit parameters are listed in Table~\ref{tab:hd107146_model}.  The debris
belt is broad, extending from 50 to 170\,AU.  The negative surface density 
power law ($p$=-0.3) is unexpected, as circumstellar disk density is generally 
expected to decrease with distance from the star.  The power law index is not 
tightly constrained, however (estimated uncorrelated uncertainty of $\pm$0.3), 
and with the relatively coarse spatial resolution, the SMA data could be 
smoothing over multiple belts of differing masses and radii.  The masses of the
two dust belts (0.41\,M$_\earth$ for the outer belt and 
1.6$\times$10$^{-4}$\,M$_\earth$ for the inner belt) are fairly typical for
multiple-belt debris systems.  
Figure~\ref{fig:hd107146} shows a comparison of the data and model in both 
the image and visibility domains, as well as for the SED.  The images 
demonstrate that the model subtracts cleanly, leaving no difference at the 
$>3\sigma$ level that would indicate an azimuthal asymmetry.  

\begin{figure*}[ht]
\epsscale{1.1}
\plotone{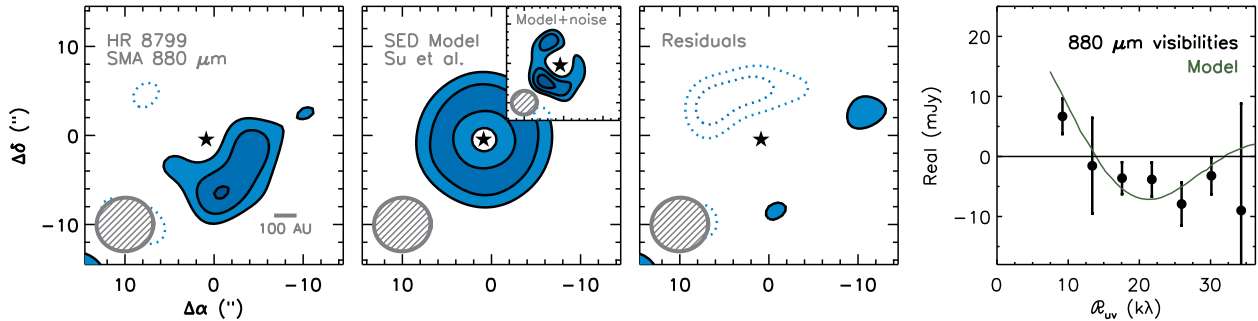}
\figcaption{
Comparison between HR~8799 millimeter data and the \citet{su09} model of the
SED.  The three image domain panels show the SMA 880\,$\mu$m data (left), 
model (center), and residuals (right), with contours 
[2,3,4]$\times$1.1\,mJy\,beam$^{-1}$ (the rms noise).  The 
6\farcs0$\times$6\farcs4 synthesized beam is indicated in the lower left of 
each panel, and the inset in the center panel shows a model prediction with
added noise at the level of the SMA data set.  It should be noted that the
arc of emission in the inset model with noise changes position depending on
the particular noise in the simulation; the arc is a common feature of all 
models, but its position angle is not.  The visibility domain plot to
the right compares the data (black points) and the model (green line). 
\label{fig:hr8799}}
\end{figure*}

\begin{table}
\caption{Best-Fit Parameters for HD~107146}
\begin{tabular}{lc}
\hline
\hline
Parameter & Value \\
\hline
$a$ ($\mu$m) & 4 \\
$\beta$ & 0.4 \\
$r_\mathrm{in}$ (AU) & 50 \\
$r_\mathrm{out}$ (AU) & 170 \\
$p$ & -0.3 \\
$M_\mathrm{disk}$ (M$_\earth$) & 0.41 \\
$M_\mathrm{belt}$ (M$_\earth$) & $1.6\times10^{-4}$ \\
\hline
\end{tabular}
\label{tab:hd107146_model}
\end{table}

\begin{figure*}[ht]
\epsscale{1.0}
\plotone{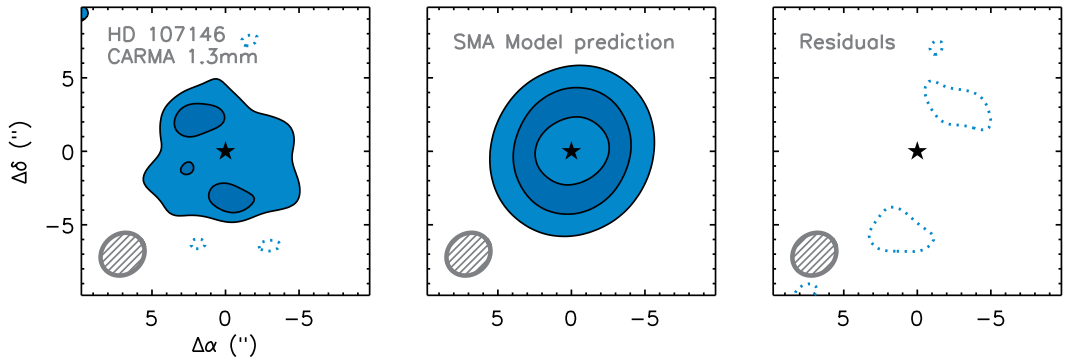}
\figcaption{An {\it a posteriori} comparison between 1.3\,mm CARMA observations of HD~107146 from \citet{cor09} and the model of disk structure based on the
SED and 880\,$\mu$m SMA visibilities.  The three panels show the CARMA data as in
Fig.~3 of \citet{cor09} (left), the model prediction at 1.3\,mm derived from
the SED and 880\,$\mu$m SMA visibilities (center), and residuals (right).  
Contours are [2,4]$\times$0.35\,mJy\,beam$^{-1}$ (the rms noise), and the 
3\farcs2$\times$2\farcs7 synthesized beam is shown in the lower left of each 
panel.  The disk structure model derived from the SED and 880\,$\mu$m SMA 
visibilities agrees well with the spatial distribution of the 1.3\,mm CARMA 
data, indicating that the emission arises from spatially coincident dust grain 
populations.
\label{fig:carma}}
\end{figure*}

The best previous spatially resolved long-wavelength imaging of the HD~107146 
disk is the 1.3\,mm CARMA observation by \citet{cor09}.  As an {\it a 
posteriori} check on the robustness of our model, we predict the spatial 
distribution of flux at the CARMA frequency from our SED + SMA model, subtract 
the model in the visibility domain using the MIRIAD task {\texttt uvmodel}, 
and image the residuals.  Figure~\ref{fig:carma} shows the results of this 
comparison: the SMA model prediction subtracts cleanly from the CARMA 
visibilities, leaving no residuals at the $>3\sigma$ level.  This also implies 
that an axisymmetric model is consistent with the millimeter-wavelength data 
obtained so far from the HD~107146 system. 

\subsection{HR~8799 Modeling Procedure}
\label{sec:hr_analysis}

Because of the very low signal-to-noise ratio in the HR~8799 data, the full 
multiparameter treatment we apply to HD~107146 is not warranted.  Instead, 
we (1) compare our data with the prediction of the SED-based model of 
\citet{su09}, and (2) explore whether the data can discriminate between a 
narrow or broad dust ring around the star.  Throughout the analysis, we 
assume that the disk is viewed face-on ($i$=0$^\circ$).  The inclination is 
likely closer to 20$^\circ$, based on a range of diagnostics including 
rotational velocity $v\sin{i}$, astrometry of the planet orbits, stability 
analyses, and the marginally resolved {\it Spitzer} observations of an 
extended halo \citep{laf09,rei09,su09,fab10,mor10}.  However, the data are
not sensitive enough to distinguish between the two inclinations; neither
are they sensitive enough to favor a particular position angle in any 
statistically meaningful way.  As described below, it is difficult to 
determine the size of the disk from the visibilities alone to within a factor 
of two.  The difference in radii derived for an inclination of 20$^\circ$ and 
0$^\circ$ is only $\sim$7\%, far less than the uncertainty in the data. 
Therefore, in the absence of a measurable  position angle it is simplest to 
assume a face-on geometry.

\citet{su09} model three components to the dust distribution in the HR 8799 
system: a warm (T$\sim$150\,K) inner dust belt, a cold (T$\sim$45\,K) outer
planetesimal belt, and an extended (R$\sim$300-1000\,AU) halo of small grains.  Since only the cold planetesimal belt contributes substantially to the flux at
millimeter wavelengths, we compare our data to this component.  The model has 
been revised to take into account the analysis of the system conducted by 
\citet{mor10}.  It includes a dust belt with constant surface density (but 
decreasing surface brightness) as a function of radius, with a sharp inner 
edge at a distance of 150\,AU from the star and an outer edge at a radius of 
300\,AU.  The location of the outer edge is consistent with constraints from
single-dish observations: since the 870\,$\mu$m flux is effectively identical 
in the 19'' APEX beam (11.2$\pm$1.5\,mJy; P.~Kalas, priv.~comm.) and the 
14'' JCMT beam \citep[10.3$\pm$1.8\,mJy][]{wil06}, the disk radius must be 
$\lesssim$300\,AU.   The total mass of the dust in the belt is 
1.2$\times$10$^{-1}$\,M$_\earth$, with grains ranging in size from 
8\,$\mu$m to 1\,mm.

The left panels of Figure~\ref{fig:hr8799} show an image-domain comparison
between the SMA data and the SED-based model of \citet{su09}, calculated at
the same wavelength as the data.  The inset in the central panel shows the 
predicted spatial distribution of emission for the \citet{su09} model observed
with the same noise level as the SMA data; the 4$\sigma$ peak offset by 6'' from
the star position and the partial arc of emission match well the observed 
morphology.  The residuals subtract fairly cleanly, with only a slight 
negative 3$\sigma$ peak.  While such a large residual could potentially point 
to an azimuthal asymmetry in the dust distribution due to the gravitational 
influence of the planet HR~8799b, which is located at a similar azimuth to the 
negative peak, the signal-to-noise ratio is too low to permit a firm 
conclusion.  The separation between the star and emission peak in the central 
panel inset is comparable to that in the data, certainly within the 
astrometric uncertainty of the observation.  The right panel of 
Figure~\ref{fig:hr8799} makes the same comparison in the visibility domain, 
where the visibilities have been azimuthally averaged to enhance the 
signal-to-noise ratio in each bin.  Within the uncertainties, the \citet{su09} 
model predicts well the spatially resolved data.
%The 
%data on baselines shorter than 20\,k$\lambda$ appear systematically fainter 
%than the model prediction, although once again the lack of signal hampers a 
%firm conclusion.  Systematically faint data at short baselines would push the 
%null in the visibility function -- the baseline lengths at which the real part 
%of the visibilities changes from positive to negative -- to shorter baselines, 
%indicating a somewhat larger ring than predicted by the SED-based model.  
%Within the uncertainties, however, the \citet{su09} model predicts well the 
%spatially resolved data.

We also perform an exploration of the SMA visibilities and SED using a 
modified version of the method described in Section~\ref{sec:hd_analysis} 
aimed at determining whether the data can distinguish between a narrow or
broad dust ring.  To do so, we fix $a$ to 1\,$\mu$m, $\beta$ to 0.5, and $p$
to 0, implying a ring of constant surface density.  We explore the narrow 
ring scenario by assuming that the width of the ring is 1\,AU, and allowing
only $r_\mathrm{in}$ to vary.  The SED is assembled from the literature, 
including 2MASS, {\it IRAS}, and {\it Hipparcos} fluxes \citep{mos89,moo06,
wil06,su09}.  To reproduce the SED, we allow $M_\mathrm{disk}$ to vary, and 
add a 150\,K belt of variable mass $M_\mathrm{belt}$, as in \citet{su09}, to 
reproduce the observed mid-IR flux.  We use a Kurucz-Lejune model photosphere
\citep{lej97} for a star of luminosity 4.92\,L$_\sun$, radius 1.4\,R$_\sun$, 
effective temperature 7250\,K, and metallicity [Fe/H]=-0.5 \citep{sad06,
mar08}.  The $\chi^2$ minimization results in a best-fit model with a ring 
radius of 170\,AU, with $M_\mathrm{disk}$ and $M_\mathrm{belt}$ equal to 
$5.5\times10^{-4}$ and $4.0\times10^{-6}$\,M$_\earth$, respectively. 

The combination of SED and visibilities can help to distinguish between a 
narrow ring and a broad belt.  The null in the visibility function of the SMA 
data occurs at roughly 12\,k$\lambda$; using Equation A11 from \citet{hug07}, 
if the underlying flux distribution were a thin ring, it would have a radius 
of about 250\,AU.  On the other hand, if it were a broad ring like the 
\citet{su09} model predicts, then using Equation A9 of \citet{hug07} the 
12\,k$\lambda$ null implies a band with a smaller inner radius by up to a 
factor of 3, depending on how quickly the flux decreases with distance from the 
star.  The 170\,AU result for the ring obtained in our SED$+$visibility 
analysis is odd in this context, since it is roughly 2/3 the radius implied 
by the visibility null and the location of the emission peak.  The mismatch 
is due to the SED fit: the $\sim$45\,K temperature implied by the SED 
\citep{su09} is warmer than would be predicted for a 250\,AU ring given the 
assumed dust properties.  The 170\,AU result also depends strongly on the choice
of a 1\,$\mu$m dust grain size.  If the grain size is allowed to vary, even
smaller grain sizes are preferred, since raising the temperature of the dust
grains results in a better match between the SED and visibilities.  It is 
therefore difficult to reconcile the large radius of a thin ring implied by 
the visibility data with the relatively warm dust implied by the SED for a
narrow ring configuration.  Yet we have already shown that the observed flux 
distribution is also consistent with that predicted for a broad band of 
emission with an inner radius of $\sim$150\,AU, as in the \citet{su09} model, 
which contains warmer dust than a narrow ring far from the star.  This is 
consistent with the location of the visibility null for a broad ring 
configuration.  Put another way, the spatially resolved SMA data {\it alone} 
are unable to discriminate between a narrow ring located $\sim$200\,AU from 
the star and a broad disk with an inner radius of $\sim$150\,AU, but the 
{\it combination} of visibilities and SED favors the latter scenario.  
Spatially resolved data with a higher signal-to-noise ratio are required to 
place more stringent constraints on the dust morphology, although the 
spatially extended \citet{su09} model is capable of reproducing both the 
detailed SED properties and the SMA data.

\section{Discussion}
\label{sec:discussion}

The emerging theoretical framework for understanding debris disk structure 
postulates belts of planetesimals that produce a range of dust grain sizes 
through a collisional cascade \citep[e.g.][and references therein]{wya08}. 
The ``birth rings'' in which dust is generated are typically assumed to be
relatively radially narrow, and millimeter-size grains are predicted to be 
confined near the birth population while smaller grains are smeared to far 
larger distances through the effects of stellar radiation \citep[e.g.][]{str06,
kuc10}.  In this context, it is interesting to compare the morphology of the 
disks derived at millimeter wavelengths to the results of spatially resolved 
imaging at optical and infrared wavelengths.  

The broad ring model derived for HD~107146 bears a striking similarity to the 
features of the scattered light observations of \citet{ard04,ert11}.  They 
infer the presence of a broad dust ring with peak optical depth at a distance 
of 130\,AU from the star and FWHM 85\,AU.  The 880\,$\mu$m SMA data have peak 
brightness at a distance of 5'' ($\sim$115\,AU) from the central star, with 
models of the spatially resolved data implying a broad ring extending from 
50 to 170\,AU, similar to the extent measured in the optical.  The negative 
surface density power law of $p\sim-0.3$ implies a near-constant surface 
brightness with radius, similar to the two-power-law behavior (with optical 
depth increasing to a radius of 130\,AU and decreasing beyond that radius) 
described for the scattered light.  However, it is not at all obvious that 
the morphology of the 880\,$\mu$m data {\it should} be so well matched to 
that of the F606W and F814W {\it HST} data, given the birth ring scenario and 
the prediction that the spatial extent of the small scatterers should exceed 
that of the millimeter-size particles.  If enough molecular gas lingers in 
the system, it is possible that even the large dust grains could remain 
entrained in gas.  It is also possible that the grain size distribution is 
concentrated at small sizes; multiwavelength imaging at higher signal-to-noise 
would be required to confirm this.

The HR~8799 disk is known to posess a large and surprisingly massive halo of
small dust grains extending between $\sim$300-1000\,AU from the star and
resolved in 70\,$\mu$m emission with {\it Spitzer} \citep{su09}.  The mass of
grains in the halo is about 15 times higher than expected for a static 
collisionally dominated disk \citep{wya07}, similar to the halo observed around
Vega.  This implies that the production of small grains is enhanced in these
systems, most likely due to heavy dynamical stirring \citep{su09}.  The 
broad ring of millimeter emission suggested by the SED model of \citet{su09}
that is consistent with the 880\,$\mu$m visibilities is likely to be generated
by the same stirring process responsible for the enhanced production of small
grains in the halo.

The other aspect of morphology worth mentioning is the degree of axisymmetry.
There is no statistically significant evidence for deviations from axisymmetry 
in the disks that might point to dynamical resonances with a large planet.  
The 3$\sigma$ negative residual in the HR~8799 observation coincides roughly 
with the position of the emission {\it peak} in the 350\,$\mu$m CSO map 
presented in \citet{pat11}, a likely indication that both features are in 
fact simply noise in the data.  We do not confirm the azimuthal asymmetry 
suggested by CSO observations in \citet{pat11} despite higher resolution and 
signal-to-noise; within the (large) uncertainties the data are so far 
consistent with a symmetric disk.  The inset in the center image panel of 
Fig.~\ref{fig:hr8799} illustrates how emission morphology in low 
signal-to-noise data sets may be misleading.  We also note that the challenge 
of interpreting faint emission morphology from debris disk has lead to 
spurious detections of clumpy structure in the Vega \citep{koe01,wil02,pie11}
and HD 107146 \citep[][this work]{cor09} debris disks.  Given these 
difficulties, greater sensitivity is necessary to point conclusively to a 
deviation from axisymmetry.  Of course, the lack of deviations from axisymmetry 
does not necessarily imply a lack of giant planets; for example, \citet{kuc10} 
predict that the dynamical influence of Neptune should be invisible at 
millimeter wavelengths in our own solar system, and that the smoothing effect 
of collisions would be enhanced for larger optical depths comparable to the 
disks observed in this paper.  

\subsection{Implications for Planetary Systems}

The location of the inner edge of the dust belt has profound implications for
our understanding of the directly-imaged planetary system orbiting HR~8799.
The visibilities alone constrain the inner radius of the dust belt to between
80 and 250\,AU, while the combination of SED and visibilities favors an inner
radius of approximately 150\,AU.  As discussed by \citet{su09}, the 90\,AU 
inner radius initially estimated from the SED is consistent with the estimated 
extent of the chaotic zone of HR~8799b.  Assuming a semimajor axis equivalent 
to the projected separation of the planet from the star (68\,AU) and the 
nominal planet mass of 5-11 M$_\mathrm{J}$ \citep{mar08}, the chaotic zone 
ranges from approximately 17 to 21\,AU from the planet depending on the 
assumed stellar age.  Assuming a nearly face-on circular orbit, this places the
chaotic zone edge at a distance of $\sim$85\,AU from the star.  It is therefore 
conceivable that the inner edge of the debris belt is being shaped by the 
outermost planet if it does indeed fall near a radius of 90\,AU.  Simulations 
by \citet{mor10} make it clear that the planetary system parameters, for 
several of the dynamical configurations described in recent stability analyses 
\citep[e.g.,][]{fab10,god09}, are consistent with an inner disk radius up to 
150\,AU in radius.  This brings it into agreement with the larger radius 
implied by the combination of SMA and SED data.  In fact, an exploration of 
the parameter space of stable planet masses and eccentricities reveals that a 
semimajor axis of 150\,AU effectively serves as an upper limit to the radius 
at which the disk is plausibly truncated by the planetary chaotic zone for any 
planetary system configuration with long-term stability.  Given the 
$\sim$20\,M$_\mathrm{J}$ upper limit on the mass of the outermost planet 
\citep{fab10}, only configurations with the outermost planet in a marginally 
stable orbit with varying semimajor axis can extend the overlapping resonances 
of the chaotic zone out to such a large distance from the star.  While a 
nonzero eccentricity of the outermost planet could also extend the radius of 
the dust belt, the dynamical stability criteria for the system generally allow 
only a planet at the lower end of the allowable mass range to achieve only a 
moderate eccentricity, ruling out the case of {\it both} a large mass and an 
eccentric orbit for HR 8799b.  More sensitive observations that pinpoint the 
location of the inner edge of the dust belt more precisely will therefore
 provide important constraints on the long-term stability of the system.  The 
similarity of the SED shape to that of Fomalhaut suggests that the inner edge 
of the disk is sharp \citep{su09}, which is also consistent with dynamical 
truncation by interactions with the planetary system.  Knowing the geometry 
of the system would aid in determining whether or not the debris belt is 
truncated by HR~8799b, since the semimajor axis of both its orbit and the 
inner edge of the debris belt could be more precisely determined.  If the 
belt is truncated by the outermost planet, this would also provide an 
independent dynamical estimate of the mass of the planet, which currently 
depends on the highly uncertain age of the system. 

No companion has to date been observed in orbit around HD~107146.  It seems 
clear that the inner edge of the debris ring is not sculpted by a stellar 
companion, as searches by \citet{met09,apa08} have ruled out the presence of
objects more massive than 10-12\,M$_\mathrm{J}$ at separations of 
$\sim$15-75\,AU from the star.  However, there are no such limits on the 
presence of a $<12$\,M$_\mathrm{J}$ companion.  \citet{cor09} discuss the 
features that a putative planet would possess if it were responsible for 
sculpting resonances in the debris disk; however, since the SMA observations 
and modeling presented here do not confirm the clumpy structure suggested by 
the CARMA data alone, the properties of a planet interacting with the debris 
would likely be somewhat different.  Assuming that the inner edge of the 
debris belt corresponds with the outer edge of the planet's chaotic zone, a 
planet with mass between 1 and 10\,M$_\mathrm{J}$ would most likely be located 
at a deprojected radial separation of $\sim$44\,AU from the star, given the 
50\,AU inner radius derived from the SMA data.

Of course, there are physical mechanisms other than dynamical interactions
with an embedded giant planet that are capable of generating the observed dust 
ring morphology.  Despite the upper limit on the gas density in the disk, even 
after most of the original mass has dissipated remnant gas can aid in shaping
broad rings and haloes in the dust \citep[e.g.,][]{tak01}.  The collisional 
cascade initiated by the formation of smaller planetesimals could also account 
for the set of broad bands implied by the observations of this system 
\citep[e.g.,][]{ken04}, and may be more consistent with the matching extent 
of the optical and millimeter emission.  However, the timescale necessary to
form large ($\sim$1000\,km) planetesimals at such a great distance from the 
star is only marginally consistent with the 160\,Myr age of the system, and
the width of the emission band is too large to arise from only one such ring
of planet formation.

\section{Summary and Conclusions}
\label{sec:summary}

We have obtained the first spatially resolved 880\,$\mu$m observations of
the dusty debris disks around the young solar analogue HD~107146 and the
multiple-planet host star HR~8799 at a resolution of 3" and 6", respectively.  
The data reveal broad bands of emission with inner radii at tens of AU from 
the central star.  Models of axisymmetric dust annuli are capable of 
reproducing both the SED and 880\,$\mu$m visibility data, with no significant 
residuals that would indicate clumpiness due to shepherding of dust grains 
into orbital resonances with planets.  The outer debris belt around HR~8799 is 
spatially resolved for the first time.  While the spatially resolved data 
alone are not of sufficiently sensitive to constrain the location of the inner 
edge of the HR~8799 disk, the SED-based model of \citet{su09} with an inner 
edge at 150\,AU from the star is shown to be consistent with the SMA data.  A 
150\,AU inner radius serves roughly as the largest radius at which the debris
belt may be plausibly truncated by the chaotic zone of HR~8799b, and more 
sensitive imaging better able to pinpoint the location of the inner edge of 
the debris belt would be extremely valuable.  The 880\,$\mu$m observations of 
HD~107146 are consistent with previous spatially resolved disk observations at 
1.3\,mm \citep{cor09} and in the near-infrared \citep{ard04}, although we 
demonstrate that the long-wavelength data are distributed smoothly to within 
the precision of the observations.  For both systems, there is some indication 
that the population of dust grains giving rise to the millimeter data are 
located closer to the star than the likely smaller populations probed by the 
short-wavelength data, consistent with theoretical predictions for collisional 
destruction leading to blow-out of small dust.  

Long-wavelength observations with higher sensitivity will be particularly 
exciting for the case of HR~8799, since they will be able to constrain the 
disk geometry (inclination and position angle) and more precisely determine 
the location of the inner edge of the dust belt, both of which will place
important dynamical constraints on the masses of the directly-imaged planets
in the system.

\acknowledgments
The authors are grateful to Stuartt Corder and John Carpenter for sharing their
HD~107146 CARMA data.  We also thank Eugene Chiang for constructive comments 
on the paper.  A.~M.~H. is supported by a fellowship from the Miller Institute 
for Basic Research in Science. 

\bibliography{ms}

\end{document}